\documentclass[11pt]{article}
\pdfoutput=1 

\usepackage{jheppub} 
\usepackage{tikz}
\usepackage{environ}
\usetikzlibrary{decorations.pathmorphing,decorations.markings,trees,shapes}
\usepackage[force]{feynmp-auto}
\def\l{\langle}
\def\r{\rangle}
\newcommand{\Dfbd}{\mathord{\buildrel{\lower3pt\hbox{$\scriptscriptstyle\leftrightarrow$}}\over {D}_{\mu}}}
\NewEnviron{scaletikzpicturetowidth}[1]{%
  \def\tikz@width{#1}%
  \begin{lrbox}{\measure@tikzpicture}%
  \BODY
  \end{lrbox}%
  \pgfmathparse{#1/\wd\measure@tikzpicture}%
  \BODY
}
\newcommand{\bea}{\begin{eqnarray}}
\newcommand{\eea}{\end{eqnarray}}

\newcommand{\vev}[1]{\langle #1 \rangle}
\newcommand{\ket}[1]{| #1 \rangle}

\usepackage{braket,slashed,bm}

\title{Renormalization Group Evolution in On-shell SMEFT}

\author[a]{Minyuan Jiang,}

\author[b]{Teng Ma,}

\author[a,c,d,e,f]{Jing Shu}
\affiliation[a]{
CAS Key Laboratory of Theoretical Physics, Institute of Theoretical Physics,
Chinese Academy of Sciences, Beijing 100190, China.}
\affiliation[b]{
Physics Department, Technion -- Israel Institute of Technology, Haifa 3200003, Israel.}
\affiliation[c]{School of Physical Sciences, University of Chinese Academy of Sciences, Beijing 100049, P. R. China.}
\affiliation[d]{CAS Center for Excellence in Particle Physics, Beijing 100049, China}
\affiliation[e]{Center for High Energy Physics, Peking University, Beijing 100871, China}
\affiliation[f]{School of Fundamental Physics and Mathematical Sciences, Hangzhou Institute for Advanced Study, University of Chinese Academy of Sciences, Hangzhou 310024, China}

\emailAdd{minyuan@itp.ac.cn}
\emailAdd{t.ma@campus.technion.ac.il}
\emailAdd{jshu@itp.ac.cn}

\abstract{We describe the on-shell method to derive the Renormalization Group (RG) evolution of Wilson coefficients of high dimensional operators at one-loop, which is a necessary part in the on-shell construction of the Standard Model Effective Field Theory (SMEFT), and exceptionally efficient based on the amplitude basis in hand. The UV divergence is obtained by first calculating the coefficients of scalar bubble integrals by unitary cuts, then subtracting the IR divergence in the massless bubbles, which can be easily read from the collinear factors we obtained for the Standard Model fields. Examples of deriving the anomalous dimensions at dimension six are presented in a pedagogical manner. We also give the results of contributions from the dimension-8 $H^4D^4$ operators to the running of $V^+V^-H^2$ operators, as well as the running of $B^+B^-H^2D^{2n}$ from $H^4D^{2n+4}$ for general $n$.}

\def\l{\langle}
\def\r{\rangle}

\begin{document}
\maketitle
\flushbottom

\section{Introduction}

The discovery of the Higgs with a mass around $125$ GeV~\cite{Chatrchyan:2012ufa, Aad:2012tfa} completes the last missing piece of the Standard Model (SM) and indicates that the SM precisely describes the fundamental interactions at lower energy scale. But with the discovery of Higgs, the Higgs naturalness problem is still mysterious and remains to be solved in the next decades. To solve this problem, the new physics (NP) should be introduced at TeV scale, such as SUSY~\cite{Wess:1974tw,Volkov:1973ix} and composite Higgs models~\cite{Kaplan:1983fs,Georgi:1984af,Dugan:1984hq}. So far the $14$ TeV LHC has not found any new physics, which may indicate that the NP scale is so high that beyond the reach of current experiment searches. With such a high NP scale, the precise measurements of the SM interactions at lower scale is an available way to search for the hints of NP, which can be well parametrized by high dimensional operators of the SM Effective Filed Theory (SMEFT). So it becomes important to understand the SMEFT for the search of NP imprints. Since the running of Wilson coefficients can significantly affect the contributions of NP to SM processes at loop level, the  anomalous dimensions of effective operators are crucial for correctly calculating the experiment observables in SMEFT without loosing any infrared informations of the NP.

Recently it was found that on-shell scattering amplitudes have remarkable advantages for the study of SMEFT, comparing with the traditional Lagrangian language. The high dimension operators can be described by unfactorizable amplitudes~\cite{Shadmi:2018xan,Ma:2019gtx}, called $amplitude$  $ basis$, without boring with the redundancies from the equation of motion and integration by part (these redundancies are automatically removed by the intrinsic properties of on-shell method: on-shell conditions and momentum conservation). With this new basis, the calculation in SMEFT can be implemented without referring to the Lagrangian. Some surprising relations and properties of EFTs, which are not manifest in quantum field theory, can be easily seen via this method. For example,  some  EFTs can be described by scattering equations~\cite{Cachazo:2013iea} uniformly or constructed/classified systematically from soft limits~\cite{ArkaniHamed:2008gz,Cheung:2016drk,Low:2014nga}. The running of Wilson coefficients of SMEFT can be strongly constrained by selection rules~\cite{Cheung:2015aba,Bern:2019wie,Jiang:2020sdh} based on unitarity cut method. 

Since the on-shell scattering amplitudes are only described by the physical degrees of freedom, the calculations in SMEFT can be very efficient via on-shell method without involving gauge fixings and ghosts. Particularly, the one-loop amplitudes can be decomposed into the sum of a basis of scalar integrals plus rational functions. And the coefficients of the scalar integrals are determined by the product of the tree-level on-shell amplitudes from the generalized unitary cuts~\cite{Bern:1994zx,Bern:1994cg,Bern:1997sc}. So the one-loop amplitudes can be obtained through simple tree-level calculations without involving any loop integrations. In the basis of scalar integrals, only the bubble integrals are UV divergent, so the anomalous dimension matrix is simply determined by the coefficients of massive and massless bubbles, which can be obtained by Stokes's Theorem~\cite{Mastrolia:2009dr} or other methods~\cite{ArkaniHamed:2008gz,Huang:2012aq,Forde:2007mi,Anastasiou:2006jv} (for massive bubble integrals) and collinear divergences of tree level amplitudes (for massless bubble integrals)~\cite{Giele:1991vf,Giele:1993dj,Kunszt:1994np}. Notice that UV divergences from massless bubble integrals are universal and only determined by renormalizable interactions so they can be directly read out without any calculations~\cite{Kunszt:1994np}. Comparing with the existing calculation of the anomalous dimension matrix of dimension six~\cite{Grojean:2013kd,Jenkins:2013wua,Alonso:2013hga,Jenkins:2013zja,Alonso:2014zka} via Feynman diagrams, the on-shell method appears to be more convenient and powerful, especially when applied in the calculations involving higher dimension operators.      

In this paper we demonstrate how to use the on-shell method to derive the anomalous dimension matrix via tree-level amplitudes and give some non-trivial examples, such as $F^3$ type operators and dimension 8 operators (the complete dimension 8 operator basis can be found in~\cite{Li:2020gnx, Murphy:2020rsh}). Since the UV divergence from a massless bubble integral is universal (only depends on the external legs attached to this bubble diagram), we list all the UV divergent factors from massless bubbles for all the SM fields. So people can directly use these results to calculate the renormalization of SMEFT operators at one-loop level without calculating this kind of UV divergences again. We find that the custodial symmetry can also explain some zeros in anomalous dimension matrices, which can not be explained by the existing selection rules. Based on unitary cuts, the anomalous dimension matrices of the operators with arbitrary dimensions that contribute to $2 \to 2$ processes can be easily expressed in universal forms and we explicitly show the universal expressions for the running of $B^+B^-H^2 D^n$ type operators generated from the insertion of general $H^4 D^{2n +4}$ type operators at one-loop level.                                 

The structure of this paper is organized as follows. A detailed discussion about the anomalous dimension calculation in SMEFT via the on-shell method is presented in Sec.~\ref{sec:Intro}, and some examples are shown in Sec.~\ref{sec:example}. The anomalous dimension matrix for dimension 8 amplitude basis $V^+ V^- H^2$ is obtained in Sec.~\ref{sec:dim8}. We show the universal results of anomalous dimensions for general amplitude basis $B^+ B^- H^2 D^{2n}$  in Sec.~\ref{sec:universal} and conclude in Sec.~\ref{sec:conclusion}. A simple example of deriving collinear divergent factor and detailed calculation of the anomalous dimension of $\mathcal{O}_{eW}$ via on-shell method are presented in App.~\ref{collinear} and App.~\ref{app:Oew}.

\section{The on-shell loop method based on unitary cut}
\label{sec:Intro}
Since the renormalization of on-shell SMEFT is induced by UV divergent part of amplitudes, in this section we explain how to derive the full UV divergences of the amplitudes via unitary cut and collinear singularities of tree-level amplitudes.  

The non-renormalizable interactions of the on-shell SMEFT can be described by the amplitude  basis 
$ \sum_i c_i \mathcal{M}_{\mathcal{O}^i} $, where $c_i$ is the Wilson coefficient. To obtain the RG equations for $c_i$, we consider the amplitude which receives tree-level contribution from $\mathcal{M}_{\mathcal{O}^i} $ as well as loop contributions with another amplitude basis $\mathcal{M}_{\mathcal{O}^j}$ insertion. The full amplitude takes the form of
\begin{align}
\mathcal{A}_i ^{\text{1-loop}} \sim c_i(\mu)-\gamma_{ij}\frac{1}{16\pi^2}c_j(\mu)(\frac{1}{2\epsilon}+\log \mu+\ldots),
\end{align}
where  the terms $\frac{1}{2\epsilon}+\log \mu$ come from the UV divergence with dimension regularization $D=4-2\epsilon$ and $\mu$ is the renormalization scale. By demanding the full amplitude being independent of the scale $\mu$, one directly obtains the renormalization group (RG) equation
\begin{align}
\frac{\text{d} c_i(\mu)}{\text{d} \log \mu}=\sum_j\frac{1}{16\pi^2}\gamma_{ij}c_j(\mu), 
\label{rge}
\end{align}
where $\gamma_{ij}$ is the the anomalous dimension matrix governing the RG running. 
\subsection{Unitarity cut and bubble coefficients}
To extract the UV divergence in the one-loop amplitude, a convenient way is to decompose it into the combination of  a basis of scalar integrals including boxes, triangles and bubbles plus rational functions~\cite{Bern:1994zx,Bern:1994cg} 
\begin{equation}
\mathcal{A}^{\text{1-loop}}=\sum_k C_4^k I_4^k+\sum_j C_3^j I_3^j+\sum_k C_2^i I_2^i+R.
\label{rge}
\end{equation}
Here the index $i$ ($j$ or $k$) labels the distinct integrals with different partition of the external legs. These integrals capture the branch cuts of the loop amplitudes and their coefficients $C^i_{4,3,2}$ can be obtained from tree level amplitudes by generalized unitary cut~\cite{Bern:2004cz,Bern:2004ky,Britto:2004nc}. The scalar bubbles are the only UV divergent integrals in four dimensions. With dimension regularization it takes the form:
\bea
I_2^i &\equiv&-i\int \frac{d^d l}{(2\pi)^d} \frac{1}{l^2(l-K)^2} 
=\frac{1}{(4\pi)^2}(\frac{1}{\epsilon}-\log \frac{-K^2}{\mu^2}+...).
\label{bubble}
\eea
So the only job to derive anomalous dimension matrix is to extract the bubble coefficients. They can be easily  obtained by using Stokes's theorem based on unitary cut~\cite{Mastrolia:2009dr}. In the following section, we will briefly discuss about this method. 

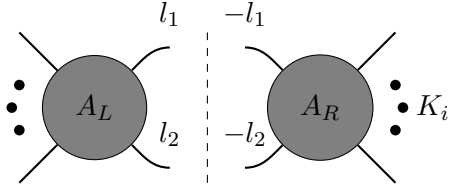
\begin{figure}
\centering
\begin{tikzpicture}

\node[draw,circle,fill=gray,inner sep=8pt] (amp1) at (0,0) {$A_L$};
\node[draw,circle,fill=gray,inner sep=8pt] (amp2) at (3,0) {$A_R$};
\fill (-1,0.3) circle (2pt);
\fill (-1.1,0) circle (2pt);
\fill (-1,-0.3) circle (2pt);
\fill (4,0.3) circle (2pt);
\fill (4.1,0) circle (2pt);
\fill (4,-0.3) circle (2pt);
\draw node at (4.5,0) {$K_i$};
\draw[thick] (amp1) to [out=45,in=180] (1,0.8) node[label=$l_1 $] {};
\draw[thick] (amp1) to [out=-45,in=180] (1,-0.8) node[label=$l_2$] {};
\draw[thick] (amp2) to [out=135,in=0] (2,0.8) node[label=$-l_1$] {};
\draw[thick] (amp2) to [out=-135,in=0] (2,-0.8) node[label=$-l_2$] {};
\draw[thick] (amp1) to [out=-135,in=45] (-1,-1) node[label=$$] {};
\draw[thick] (amp1) to [out=135,in=-45] (-1,1) node[label=$$] {};
\draw[thick] (amp2) to [out=45,in=-135] (4,1) node[label=$$] {};
\draw[thick] (amp2) to [out=-45,in=135] (4,-1) node[label=$$] {};
\draw[dashed] (1.5,-1) -- (1.5,1);
\end{tikzpicture}
\caption{$K_i$-channel double cut.} \label{fig:cut}
\end{figure}

\subsection{Extraction of bubble coefficient}
Since there are two propagators in the loop of bubble integrals,  the bubble coefficients can be extracted by double cuts.  
To extract the coefficient $C^i_2$, the  $K_i$-channel  double cut should be implemented  to $\mathcal{A}^{\text{1-loop}}$ in eq.~\ref{rge}, as illustrated in Fig.~\ref{fig:cut}. The left hand side of this equation  becomes
\bea
\text{Cut}_{K_i}[\mathcal{A}^{\text{1-loop}}]= \int d \text{LIPS}_i  \sum_{h_i} \mathcal{A}_L^{\text{tree}} (l_1, l_2,h_i) \mathcal{A}_R^{\text{tree}}(-l_1,-l_2, -h_i),  
\label{eq:disc_l}
\eea
where $d \text{LIPS}_i=d l_1^4 d^4 l_2^4 \delta^{(+)}(l_1^2)\delta^{(+)}(l_2^2)\delta(l_1+l_2-K_i)$ is the Lorentz-invariant phase space associated with the $K_i$-channel cut, $l_{1,2}^\mu$ is momentum of the propagators, $\mathcal{A}_L^{\text{tree}}, \mathcal{A}_R^{\text{tree}}$ are tree level amplitudes on each side of the cut and $h_i$ is the polarization configuration of the cutted internal legs.

The coefficient of bubble is proportional to the rational terms of the double cut, because the un-cutted propagators in the $I_{3,4}$ will make the integration variable $l_{1}$ appear in the denominators and thus contribute only irrational terms.  So the bubble coefficients is given by
\bea
 C_2^i= -\frac{1}{2\pi i} \text{Rational} [\int d \text{LIPS}_i  \sum_{h_i} \mathcal{A}_L^{\text{tree}} \mathcal{A}_R^{\text{tree}} ],
\label{eq:c2}
\eea
where the factor $-2\pi i$ is from the double cut  on bubble integral $I_2^i$ (will be seen in the following discussion). 
To efficiently calculate the phase space integral,  the loop momentum can be parametrized as  
\bea
(l_1)_{a \dot{a}}=t \lambda_a \tilde{\lambda}_{\dot{a}},
\eea
so that the phase space integral can be written as 
\bea
\int d \text{LIPS}_i=\int t dt \int_{\bar{\lambda}=\tilde{\lambda}} \frac{ \l \lambda d\lambda \r [\tilde{\lambda} d\tilde{\lambda}]}{\l l| K_i |l]} \delta(t-\frac{K^2_i}{\l l |K_i| l]}).
\eea

Notice that we are integrating over the contour $\bar{\lambda}=\tilde{\lambda}$ so that the loop momentum is real. The spinor variables $\lambda (\tilde{\lambda})$ can be further decomposed  into a basis of two massless spinors with a complex coefficient $z$, 
\bea
\ket{ \lambda} = \ket{p}+z \ket{q}, |\lambda ]= |p]+\bar{z} |q],
\eea
where $p_\mu$ and $q_\mu$ are two null momenta satisfying
$K_\mu^i=p_\mu+q_\mu.$
With this parametrization, the phase space integral can be expressed in terms of the integration of complex variable $z$,
\bea
\int d \text{LIPS}_i= \oint dz \int d \bar{z} \int  dt  \;t^2 \delta(t-\frac{1}{(1+z\bar{z})}).
\eea
Then the cut of the bubble integral can be easily evaluated and equal to
\bea
\text{Cut} [I_2^i]&=&\oint dz \int d \bar{z} \int t^2 dt \delta(t-\frac{1}{(1+z\bar{z})})
=\int dz \frac{-1}{(1+z\bar{z})z}
=-2\pi i.
\eea
In the last two step, we first integrate over $\bar{z}$ and then sum over the residues at all poles of $z$. This result explains the $-2 \pi i$ factor in Eq.~\ref{eq:c2}.  Under this parametrization the cutted amplitude becomes
\bea 
\Delta_i(z,\bar z) &\equiv&   \int d \text{LIPS}_i  \sum_{h_i} \mathcal{A}_L^{\text{tree}} \mathcal{A}_R^{\text{tree}} 
= \oint dz \int d \bar z   t^2 \sum_{h_i}  \mathcal{A}_L(t, z, \bar z ) \mathcal{A}_R (t,z, \bar z )\vert_{t= \frac{1}{1+z \bar z}}.
\eea
 After perform the $z$-integration via Cauchy's Residue Theorem, finally  $C_2^i$ in  Eq.~\ref{eq:c2} can be easily evaluted via the following expression,
 \bea
 C_2 = \frac{\Delta_i ^{ \text{Rational}}}{-2\pi i} =-\text{Res}_{z=0} F^{\text{Rational}}(z,\bar z)  -\text{Res}_{z \ne 0} F^{\text{Rational}}(z,\bar z),
 \eea 
 where  $F^{\text{Rational}} $ is the rational part of $F(z,\bar z) =\int d \bar z  t^2\sum_{h_i}  \mathcal{A}_L(t, z, \bar z ) \mathcal{A}_R (t,z, \bar z )\vert_{t= \frac{1}{1+z \bar z}}$. 
\subsection{Collinear divergence}
Using unitary cut we can not get the full UV divergences of loop amplitudes. Only the coefficients of massive bubble ($K^2_i>0$) can be obtained via unitary cut. Massless bubbles with $K^2_i=0$ also contain UV divergences, which simply vanish in dimension regularization due to the cancellation between UV and collinear IR divergence. Since the physical cross section is free of collinear divergences, the collinear divergences of the tree amplitudes must be cancelled by collinear loop IR divergences. So the UV divergence in massless bubbles can be extracted by calculating collinear divergences of the corresponding tree amplitudes.  
The one-loop collinear IR divergence can be parametrized as~\cite{Giele:1991vf,Giele:1993dj,  Kunszt:1994np} 
\bea
\mathcal{A}_{\text{n,col}}^{\text{1-loop}} = -(\frac{1}{4\pi} )^2 \sum_{a}^n \frac{\gamma(a)}{\epsilon} \mathcal{A}^{\text{tree}},
\eea
where the sum is over all external legs and $\gamma(a)$ is the collinear factor for each particle $a$.
We want to emphasize that the collinear factors only depend on the external legs and are universal for SMEFT. In Appendix~\ref{collinear} we show one example of deriving the collinear factors. For all the SM fields the collinear factors are obtained to be,
\bea \label{eq:collinear}
\gamma(H^a)&=& \gamma(H^{\dagger \dot a}) =2y_h^2g_1^ 2  -\frac{1}{2} \text{Tr}[ N_c Y_u^\dagger Y_u+ N_c Y_d^\dagger Y_d+Y_e^\dagger Y_e] +2g_2^2C_2(2), \nonumber \\
\gamma(B) &=& -\frac{g_1^{2}}{3} \Big[ n_g N_c ( y_{q}^2 +y_{u}^2 +y_{d}^2) + n_{g}( y_{\ell}^2 +y_{e}^2 )+y_h^2 \Big], \nonumber \\
\gamma(W^{a}) &=& g_2^2\Big[\frac{11}{3} -\frac{1}{3} ( \frac{n_g}{2}N_c +\frac{n_{g}}{2} +\frac{1}{4}) \Big], \nonumber \\
\gamma(g)&=&   g_3^2 \Big( \frac{11N_c}{6} -\frac{1}{3} n_g \Big), \nonumber \\
\gamma(\ell) &=& \frac{3}{2} g_2^2 C_2(2) + \frac{3}{2} y_{\ell}^2 g_1^{ 2} -\frac{1}{4} Y_e^\dagger Y_e, \nonumber \\
\gamma(e) &=& \frac{3}{2} y_{e}^2 g_1^{2}   -\frac{1}{2}Y_e^\dagger Y_e, \nonumber \\
\gamma(q) &=& \frac{3}{2} g_3^2 C_2(N_c)+\frac{3}{2} g_2^2 C_2(2)+ \frac{3}{2} y_{q}^2 g_1^{ 2} - \frac{1}{4} (Y_u^\dagger Y_u+ Y_d^\dagger Y_d),  \nonumber \\
\gamma(u) &=& \frac{3}{2} g_3 C_2(N_c)+ \frac{3}{2} y_{u}^2 g_1^{ 2}  - \frac{1}{2}Y_u^\dagger Y_u,\nonumber \\
\gamma(d) &=& \frac{3}{2} g_3 C_2(N_c)+ \frac{3}{2} y_{u}^2 g_1^{ 2}  - \frac{1}{2} Y_d^\dagger Y_d.
\eea
Here $q$ and $l$ are $SU(2)$ doublets for left hand quarks and leptons, while $u,d$ and $e$ are right hand singlets. $g_3$, $g_2$ and $g_1$ are the gauge couplings of $SU(3)_C \times SU(2)_L \times U(1)_Y$, with $y_i$ being the hypercharge. $Y_{u}$ , $Y_{d}$ and $Y_e$ are Yukawa couplings. $N_c=3$ is QCD color number; $n_g=3$ is the number of  generations. And $C_2(N)=\frac{N^2-1}{2N}$.

So for any SMEFT loop calculation, the UV divergences of massless bubbles can be directly read from Eq.~\ref{eq:collinear}  without  any calculation.

Combining the UV divergences from massive bubbles with collinear IR divergences, we can finally obtain the anomalous dimension matrix correctly. In the next section we will give some non-trivial examples to clearly show how  to get the anomalous dimension matrix systematically via on-shell method.

\section{Examples for calculation of anomalous dimension matrix}
\label{sec:example}
In this section we give some examples to demonstrate the on-shell loop method for calculating the RG running of SMEFT in detail.   
\subsection{$\mathcal{O}_{HB}$}
\tikzset{
    photon/.style={decorate, decoration={snake}},
    electron/.style={draw=blue, postaction={decorate},
        decoration={markings,mark=at position .55 with {\arrow[draw=blue]{>}}}},
    gluon/.style={decorate, draw=magenta,
        decoration={coil,amplitude=4pt, segment length=5pt}} 
}
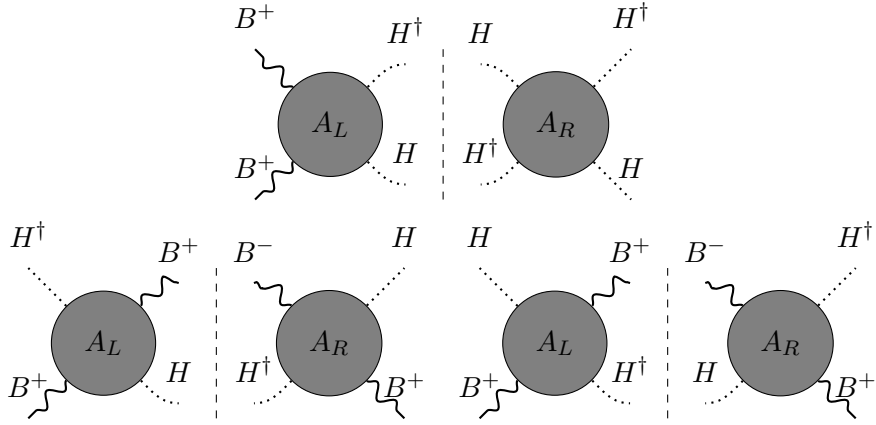
\begin{figure}
\centering
\begin{tikzpicture}
\node[draw,circle,fill=gray,inner sep=8pt] (amp1) at (0,0) {$A_L$};
\node[draw,circle,fill=gray,inner sep=8pt] (amp2) at (3,0) {$A_R$};

\draw[thick,dotted] (amp1) to [out=45,in=180] (1,0.8) node[label=$H^\dagger $] {};
\draw[thick,dotted] (amp1) to [out=-45,in=180] (1,-0.8) node[label=$H$] {};
\draw[thick,dotted] (amp2) to [out=135,in=0] (2,0.8) node[label=$H$] {};
\draw[thick,dotted] (amp2) to [out=-135,in=0] (2,-0.8) node[label=$H^\dagger$] {};
\draw[thick,photon] (amp1) to [out=-135,in=45] (-1,-1) node[label=$B^+$] {};
\draw[thick,photon] (amp1) to [out=135,in=-45] (-1,1) node[label=$B^+$] {};
\draw[thick,dotted] (amp2) to [out=45,in=-135] (4,1) node[label=$H^\dagger$] {};
\draw[thick,dotted] (amp2) to [out=-45,in=135] (4,-1) node[label=$H$] {};
\draw[dashed] (1.5,-1) -- (1.5,1);
\end{tikzpicture}

\begin{tikzpicture}
\node[draw,circle,fill=gray,inner sep=8pt] (amp1) at (0,0) {$A_L$};
\node[draw,circle,fill=gray,inner sep=8pt] (amp2) at (3,0) {$A_R$};

\draw[thick,dotted] (amp1) to [out=-45,in=180] (1,-0.8) node[label=$H$] {};
\draw[thick,dotted] (amp1) to [out=135,in=-45] (-1,1) node[label=$H^\dagger$] {};
\draw[thick,photon] (amp2) to [out=135,in=0] (2,0.8) node[label=$B^-$] {};
\draw[thick,dotted] (amp2) to [out=-135,in=0] (2,-0.8) node[label=$H^\dagger$] {};
\draw[thick,photon] (amp1) to [out=-135,in=45] (-1,-1) node[label=$B^+$] {};
\draw[thick,photon] (amp1) to [out=45,in=180] (1,0.8) node[label=$B^+$] {};
\draw[thick,dotted] (amp2) to [out=45,in=-135] (4,1) node[label=$H$] {};
\draw[thick,photon] (amp2) to [out=-45,in=135] (4,-1) node[label=$B^+$] {};
\draw[dashed] (1.5,-1) -- (1.5,1);
  \begin{scope}[xshift=6cm]
  \node[draw,circle,fill=gray,inner sep=8pt] (amp1) at (0,0) {$A_L$};
  \node[draw,circle,fill=gray,inner sep=8pt] (amp2) at (3,0) {$A_R$};

\draw[thick,dotted] (amp1) to [out=-45,in=180] (1,-0.8) node[label=$H^\dagger$] {};
\draw[thick,dotted] (amp1) to [out=135,in=-45] (-1,1) node[label=$H$] {};
\draw[thick,photon] (amp2) to [out=135,in=0] (2,0.8) node[label=$B^-$] {};
\draw[thick,dotted] (amp2) to [out=-135,in=0] (2,-0.8) node[label=$H$] {};
\draw[thick,photon] (amp1) to [out=-135,in=45] (-1,-1) node[label=$B^+$] {};
\draw[thick,photon] (amp1) to [out=45,in=180] (1,0.8) node[label=$B^+$] {};
\draw[thick,dotted] (amp2) to [out=45,in=-135] (4,1) node[label=$H^\dagger$] {};
\draw[thick,photon] (amp2) to [out=-45,in=135] (4,-1) node[label=$B^+$] {};
\draw[dashed] (1.5,-1) -- (1.5,1);
    \end{scope}
\end{tikzpicture}
\caption{Unitary cuts in calculating the running of $O_{HB}$. The upper plot is the $s$ channel cut and the lower two plots are $t$ channel cuts.} \label{fig:OHBcut}
\end{figure}

We first focus on a simplest case: the contributions proportional to $U(1)_Y$ gauge interactions to the running of the dimension 6 operator $\mathcal{O}_{HB}=H^\dagger H B^{\mu \nu}B_{\mu \nu}$.  In the amplitude basis, this operator corresponds to the local amplitude
\bea
\mathcal{A}(B^+,B^+,H^\alpha,H^{\dagger \dot \beta})=2C_{HB}\delta^{\alpha \dot \beta}[12]^2.
\label{B+B+}
\eea
Here the superscript "$+$" denotes the positive helicity.

The Higgs $U(1)_Y$ gauge  interactions and quartic term can be expressed as 
\bea
\mathcal{L} &=& -\frac{1}{4}B_{\mu \nu} B^{\mu \nu} +|D_\mu H|^2 -\frac{1}{4}\lambda |H|^4.
\label{U1L}
\eea
where $D_\mu =\partial_\mu -ig_1 y_h B_\mu$.

Now let's consider the amplitude   $\mathcal{A}(B^+,B^+,H^{\dagger \dot \alpha},H^\alpha)$ at one-loop. As shown in Fig.~\ref{fig:OHBcut}, applying different double cuts, this amplitude are separated into dimension 6 part and dimension 4 part. The dimension 6 part can be read from Eq.~\ref{B+B+} and the dimension 4 part can be derived from Eq.~\ref{U1L},
\bea
\mathcal{A}^4(H^\alpha H^{\dagger \dot \alpha} B^{+} B^-) &=&  -2y_h^2g_1^2 \delta^{\alpha \dot \alpha} \frac{\vev{14} \vev{24}}{\vev{13} \vev{23}}, \\\label{HHBB}
\mathcal{A}^4(H^\alpha H^{\dagger \dot \alpha} H^\beta  H^{\dagger \dot \beta}) &=&(\delta^{\alpha \dot \alpha} \delta^{\beta \dot \beta} +\delta^{\alpha \dot \beta} \delta^{\beta \dot \alpha} )(  -\frac{\lambda}{2} +y_h^2g_1^2)+2y_h^2g_1^2\Big[\delta^{\alpha \dot \alpha} \delta^{\beta \dot \beta} \frac{\vev{24}[24] }{\vev{12}[12]}+\delta^{\alpha \dot \beta} \delta^{\beta \dot \alpha} \frac{\vev{24}[24] }{\vev{14}[14]}  \Big].\nonumber\\\label{HHHH}
\eea

First consider the $s$-channel cut (the upper plot in Fig.~\ref{fig:OHBcut}). With the tree level amplitudes Eq.~\ref{B+B+} and Eq.~\ref{HHHH} it can be expressed as
\bea
&&\text{Cut}_{12}[\mathcal{A}(B^+ (p_1)B^+(p_2) H^{\dagger \dot \alpha}(p_3) H^\alpha)(p_4)]\nonumber\\
&=&\int d \text{LIPS} \mathcal{A}_L(B^+(p_1)B^+(p_2) H^{\beta}(l_1) H^{\dagger \dot \beta} (l_2))\delta^{\beta \dot \sigma} \delta^{\sigma \dot \beta}  \mathcal{A}_R(H^{\dagger \dot\sigma} (-l_1)H^\sigma (-l_2) H^{\dagger \dot \alpha} (p_3)H^\alpha(p_4) ) \nonumber\\
&=&\int d \text{LIPS} C_{HB} \delta^{\sigma \dot \sigma  } [12]^2\Big[(\delta^{\sigma \dot \sigma} \delta^{\alpha \dot \alpha} +\delta^{\sigma \dot \alpha} \delta^{\alpha \dot \sigma} )(  -\frac{\lambda}{2} +y_h^2g_1^2) \nonumber\\
&&+2y_h^2g_1^2\Big(\delta^{\sigma \dot \sigma} \delta^{\alpha \dot \alpha} \frac{\vev{-l_2 3}[-l_2 3] }{\vev{-l_1-l_2}[-l_1-l_2]}+\delta^{\sigma \dot \alpha} \delta^{\alpha \dot \sigma}  \frac{\vev{-l_2 3}[-l_2 3] }{\vev{-l_1 3}[-l_1 3]}  \Big)\Big],
\eea
where $l_{1,2}$ is the momentum of the cutted internal Higgs leg.
We then use the relations  $|-l\r=i| l\r,|-l]=i| l]$~\footnote{For  internal fermion leg, the complex factor $i$ can be removed because the fermion statistic cancels the minus sign from momentum flipping.}  and $l_1+l_2=p_3+p_4$ to express it as function of $l_1$. Following  the procedure presented in the previous section, we should parametrize the loop momentum as
\bea \label{eq:para_12}
|l_1\r =\sqrt{t}( \ket{3} +z \ket{4}) \quad |l_1] =\sqrt{t}( |3] +\bar z |4]),
\eea
and then the bubble coefficient can be extracted through Stokes' Theorem, 
\bea
\Delta_{12} &=&[12]^2\delta^{\alpha \dot \alpha}\oint dz \int d \bar z t^2\Big( (-\frac{3}{2}\lambda+3y_h^2g_1^2)+2y_h^2g_1^2(-2t^3+\frac{1}{z \bar z}t^2)  \Big)|_{t =(1+z \bar z)^{-1} }\nonumber\\
&=& [12]^2\delta^{\alpha \dot \alpha}\oint dz \Big( (-\frac{3}{2}\lambda+3y_h^2g_1^2)\frac{-1}{z(1+z \bar z)}+2y_h^2g_1^2(\frac{1}{z (1+z \bar z)^2}+\frac{1}{z(1+z \bar z)}+\frac{\log(-z \bar z)-\log(1+z \bar z)}{z})  \Big)\nonumber\\
&=&2\pi i [12]^2\delta^{\alpha \dot \alpha} (\frac{3}{2}\lambda+y_h^2g_1^2).
\eea
In the last step we discard the log terms, take the residue at $z=0$ and set $\bar z=z$. Finally we can get the bubble coefficient from s-channel double cut 
\bea
C_2^{12} =-(y_h^2g_1^2 +\frac{3\lambda}{2} )C_{HB} \delta^{\alpha \dot \alpha}[12]^2.
\eea

Following the same procedure, The contribution from t-channel cut is given by
\bea \label{eq:cut}
&&\text{Cut}_{13}[\mathcal{A}(B^+(p_1) B^+(p_2) H^{\dagger \dot \alpha}(p_3) H^\alpha)(p_4)]\nonumber\\ &=& \mathcal{A}_L(B^+(p_1)H^{\dagger \dot \alpha}(p_3)B^+(l_1)H^\beta (l_2))\delta^{\beta \dot \beta} \mathcal{A}_R(B^-(-l_1)H^{\dagger \dot \beta}(-l_2)B^+(p_2) H^\alpha(p_3) ) \nonumber\\
&+& \mathcal{A}_L(B^+(p_2)H^\alpha(p_4) B^+(-l_1) H^{\dagger \dot \beta}(-l_2))\delta^{\beta \dot \beta} \mathcal{A}_R(B^-(-l_1)H^{ \beta}(-l_2)B^+(p_1) H^{\dagger \dot \alpha}(p_3) ) \nonumber\\
&=& -2y_h^2g_1^2 C_{HB} \delta^{\alpha \dot \alpha} [1l_1]^2\frac{\vev{4 -l_1} \vev{-l_2-l_1}}{\vev{42} \vev{-\ell_2 2}} +(1 \leftrightarrow 2,3 \leftrightarrow 4). 
\eea
 Using Stokes' Theorem again after proper parametrization, we can get the bubble coefficient for t-channel cut,
\bea
C_2^{13} =-2y_h^2g_1^2 C_{HB} \delta^{\alpha \dot \alpha} [12]^2.
\eea
The u-channel cut is the same as t-channel cut under $p_1 \leftrightarrow p_2$ and we get 
\bea
C_2^{13} =C_2^{14}. 
\eea
So the total contribution from massive bubble integral is
\bea
C_2 = C_2^{13} +C_2^{14}  +C_2^{12}  = -(5y_h^2g_1^2 +\frac{3\lambda}{2} )C_{HB} \delta^{\alpha \dot \alpha}[12]^2.
\eea

The parts of UV divergences cancelled by IR divergences can be read from Eq.~\ref{eq:collinear} according to the external legs:
\bea
C_{IR} =- (2\gamma(B) +2 \gamma(H))  \delta^{a \dot a}C_{HB}[12]^2.   
\eea    
So we can find the total UV divergences of operator $\mathcal{O}_{HB}$ from $U(1)_Y$ gauge interactions at one-loop level as
\bea
C_{UV} &=& C_2 -C_{IR} =- \Big[y_h^2g_1^2 \big(5-2\gamma(H) -2\gamma(B) \big) +\frac{3\lambda}{2} \Big]\frac{1}{(4\pi)^2} \delta^{a \dot a} C_{HB}[12]^2  \nonumber \\
 &=& - \Big[\frac{5}{3}y_h^2g_1^2 +\frac{3\lambda}{2} \Big]\frac{1}{(4\pi)^2} \delta^{a \dot a} C_{HB}[12]^2.
\eea
 So the running of $C_{HB} $ is
\bea
\dot{C}_{HB}  = 2\Big[\frac{5}{3}y_h^2g_1^2 +\frac{3\lambda}{2} \Big] C_{HB},
\eea
where $\dot{C}_{HB}\equiv (4\pi)^2 \mu \frac{d C_{HB}}{d \mu}$. This expression is exactly the same as the results in~\cite{Alonso:2013hga,Jenkins:2013zja}.

\subsection{$F^3$ type operators}
For the $F^3$ type operators, with $F$ denoting the field strength of non-Abelien gauge fields, the leading amplitude is three-point scattering amplitude, which does not depend on any UV scale if the three external legs are on-shell. So its UV divergences can not be extracted through unitarity cut. However we can calculate its next leading $4$-points amplitude to derive its RG running.       
The color-ordered leading amplitude of dimension 6 operator $\mathcal{O}_{G^3} =C_{G^3} \text{Tr}[G^3_{\mu \nu}]$ can be expressed as
\bea
\mathcal{A}(g^+_1 g^+_2 g^+_3)=C_{G^3} [12][23][31].
\eea
It contributes to the 4-point all plus tree-level amplitude, which can be constructed through all-line shift~\cite{Cohen:2010mi},
\bea
\mathcal{A}^{\text{tree}}(g^+_1g^+_2g^+_3g^+_4)= 2  g_3 C_{G^3} \frac{[12][13][42]}{\vev{34}}.
\eea
According to unitary cut, it is easy to find that the loops with quark internal legs does not contain UV divergences, because the the helicity selection rules forbid the tree level amplitude for two quarks and two $g^+$ at dimension 4. So only $\mathcal{O}_{G^3}$ insertion contribute to itself  RG running.          
Following the same procedures as above, we can get the coefficients of the bubble integral,
\bea
C_{2} = -6g^2_3 N_c \mathcal{A}^{\text{tree}}(g^+_1g^+_2g^+_3g^+_4).
\eea
After including the massless bubble contributions which can be directly read from Eq.~\ref{eq:collinear}, the total divergences in the one-loop amplitude is
\bea
C_{UV} =\frac{(-6g_3^2 N_c + 4 g_3^2 \gamma(g))}{(4\pi)^2}\mathcal{A}^{\text{tree}}(g^+_1g^+_2g^+_3g^+_4).
\eea

These divergences of this four-point amplitude contain both the gauge coupling $g_s$ and $C_{G^3}$ renormalization and with requirement that the amplitude independent on the renormalizable scale  we can get the following RG equation
\bea \label{eq:RG_EQ}
(4\pi)^2 \beta_{g_3} C_{G^3} + g_3 \dot{C}_{G^3} =4g_3^3(3 N_c - 2 \gamma(g)) C_{G^3},
\eea
where $\beta_{g_3}$ is the beta function of gauge coupling $g_3$. We can also derive $\beta_{g_3}$ through on-shell method,
\bea
\beta_{g_3} =-\frac{2g^3_3}{(4\pi)^2} \gamma(g).
\eea
Substituting the expression of $\beta_{g_3}$ into the Eq.~\ref{eq:RG_EQ}, we can get the RG running of $C_{G^3}$ 
\bea
\dot{C}_{G^3}= \big[12 N_c -6\gamma(g)\big]g^2_3C_{G^3}.
\eea
With these examples we can find that on-shell method is very efficient to calculate SMEFT RG running. We do not need to do loop calculation. All the divergences can be extracted from the tree level amplitudes.     
We also present  more complicated calculations for the operator $\mathcal{O}_{eW}$  in App.~\ref{app:Oew}.  

\section{Anomalous dimensions at dimension 8: the $V^+V^-H^2$ example}
\label{sec:dim8}
The method introduced above can be applied to efficiently obtain the RG running of higher dimension operators. In this section we consider the running at dimension 8. In particular we present the contribution from the $H^4$ type local amplitudes to the RG running of the coefficients of the $V^+V^-H^2$ type amplitude.

The $V^+V^-H^2$ amplitude at dimension 8 is important in phenomenology because it gives leading BSM correction to the $VVHH$ scattering in the SMEFT, due to the non-interference at dimension 6~\cite{Azatov:2016sqh}. Also this amplitude basis is only generated at one-loop order when we integrate out some heavy particles in a weakly coupled UV theory~\cite{Jiang:2020sdh,Craig:2019wmo,Arzt:1993gz, Giudice:2007fh,Einhorn:2013kja}, which makes the contribution from the mixing with a potentially tree level generated local amplitude (operator) important. So in this section we calculate the RG running of this coupling from the mixing with $H^4$ amplitude basis at dimension 8. Notice that if we ignore the fermions, this is the only leading contribution. All others, including the contributions from the loops containing two dimension 6 amplitude basis, are more than one-loop suppressed if we take into account the tree/loop classification of the local amplitude. 

There are three independent $H^4$ type local amplitudes at dimension 8. It is convenient to write them in the following form:
\bea
\mathcal{A}(H^\alpha  H^\beta  H^{\dagger \dot \alpha} H^{\dagger \dot \beta})_{\text{dim8}} &\supset&  T^+_{ \alpha\beta\dot\alpha\dot\beta} C_{0,2}^{H^4+} (s_{13} -s_{23})^{2}, \; \; T^+_{\alpha\beta\dot\alpha\dot\beta} C_{2,0}^{H^4+}s_{12}^2, \; \; T^-_{\alpha\beta\dot\alpha\dot\beta} C_{1,1}^{H^4-} s_{12} (s_{13} -s_{23}), \nonumber\\
\label{h4}
\eea
where $T^\pm _{\alpha\beta\dot\alpha\dot\beta} \equiv \delta_{\alpha \dot\alpha} \delta_{\beta \dot\beta}  \pm \delta_{\beta \dot\alpha} \delta_{\alpha \dot\beta} $. 

The dimension 8 $V^+V^-H^2$ type amplitudes can be written as:
\bea \label{eq:HV8}
\mathcal{A}(B^+ B^- H^\alpha H^{\dagger \dot \beta} )_{\text{dim8}} &=&C_{H^2B^+B^-}\delta_{\alpha\dot \beta}[1|3|2\r^2,\nonumber \\
\mathcal{A}( W^{a+} W^{b-} H^\alpha H^{\dagger \dot \beta})_{\text{dim8}} &\supset& C_{H^2W^+W^-}^+T^{ab+}_{\alpha \dot \beta}[1|3|2\r^2,\; \; C_{H^2W^+W^-}^-T^{ab-}_{\alpha \dot \beta}[1|3|2\r^2,\nonumber \\
\mathcal{A}(W^{a+} B^- H^\alpha H^{\dagger \dot \beta} )_{\text{dim8}} &=&C_{H^2W^+B^-}\tau^a_{\dot \beta \alpha}[1|3|2\r^2,\nonumber \\
\mathcal{A}(B^{+} W^{a-} H^\alpha H^{\dagger \dot \beta} )_{\text{dim8}} &=&C_{H^2B^+W^-}\tau^a_{\dot \beta \alpha}[1|3|2\r^2,
\eea
where $\tau^a_{\alpha \dot \beta}$ is Pauli matrix, $T^{ab+}_{\alpha \dot \beta} =\delta^{ab}\delta_{\alpha \dot \beta}$ and $T^{ab-}_{\alpha \dot \beta} =i \epsilon^{abc} \tau^c_{\alpha \dot \beta}$.

The loop contributions to the running can be obtained by gluing the $H^4$ amplitude basis in Eq.~\ref{h4} with the $V^+V^-H^2$ amplitudes in SM, which are
\bea
\mathcal{A}^{\text{SM}}(H^\beta,H^{\dagger \dot \alpha},B^+,W^{i-})&=&-g_1g_2y_h(\tau^i)^{\dot \alpha} _\beta \frac{\vev{14} \vev{24}}{\vev{13} \vev{23}},\\
\mathcal{A}^{\text{SM}}(H^\beta,H^{\dagger \dot \alpha},W^+,B^{i-})&=&-g_1g_2y_h(\tau^i)^{\dot \alpha} _\beta \frac{\vev{14} \vev{24}}{\vev{13} \vev{23}},\\
\mathcal{A}^{\text{SM}}(H^{\beta},H^{\dagger \dot \alpha}, W^{a+},W^{b-})&=&-2g_2^2\frac{\l4|1|3]^2}{\l34\r[34]}\big(\frac{(t^b t^a)^{\dot \alpha}_\beta}{\l24\r[24]}+\frac{(t^a t^b)^{\dot \alpha}_\beta}{\l23\r[23]}\big),
\eea
in addition to the $B^+B^-H^\dagger H$ amplitude in Eq.~\ref{HHBB}. Here $t^a=\tau^a/2$ are the $SU(2)$ generators.

Applying the unitary method, we obtain the RG running of $V^+ V^- H^2$ type operators as following:
\bea
\dot C_{H^2B^+B^-}&=&-2g_1^2 y_h^2 (C_{0,2}^{H^4+}+C_{2,0}^{H^4+}+\frac{1}{3}C_{1,1}^{H^4-}),\nonumber \\
\dot C_{H^2B^+W^-}&=&-\frac{1}{3}g_1g_2 y_h (C_{0,2}^{H^4+}+C_{2,0}^{H^4+}-C_{1,1}^{H^4-}),\nonumber \\
\dot C_{H^2W^+B^-}&=&-\frac{1}{3}g_1g_2 y_h (C_{0,2}^{H^4+}+C_{2,0}^{H^4+}-C_{1,1}^{H^4-}),\nonumber \\
\dot C_{H^2W^+W^-}^+&=&-\frac{1}{2}g_2^2 (C_{0,2}^{H^4+}+C_{2,0}^{H^4+}+\frac{1}{3}C_{1,1}^{H^4-}),\nonumber \\
\dot C_{H^2W^+W^-}^-&=&0.
\eea
The dependence of these equations on the specific combinations $(C_{0,2}^{H^4+}+C_{2,0}^{H^4+}+\frac{1}{3}C_{1,1}^{H^4-})$ and $(C_{0,2}^{H^4+}+C_{2,0}^{H^4+}+\frac{1}{3}C_{1,1}^{H^4-})$ are expected by demanding the right angular momentum and $SU(2)_L$ quantum number\cite{Jiang:2020sdh}. The zero in the last equation can be understood from the fact that the there is $SO(4)$ custodial symmetry in both  $H^4$ operators and $\mathcal{A}^{SM}(H^{\beta}H^{\dagger \dot\alpha} W^{a+}W^{b-})$, while dimension $8$ amplitude basis  $\mathcal{A}( W^{a+} W^{b-} H^\alpha H^{\dagger \dot \beta})_{\text{dim8}}=C_{H^2W^+W^-}^-T^{ab-}_{\alpha \dot \beta}[1|3|2\r^2$ violate this symmetry. So custodial symmetry can provide some new selection rules, which can not be explained by existing selection rules, based on on-shell method. 

\section{Universal results for anomalous dimensions}
\label{sec:universal}
From above examples, we find that the anomalous dimension matrix is strongly dependent on the external legs of amplitude basis. With on-shell method, the structure of anomalous dimension matrix can be clearly seen and some universal results for the RG running of the amplitude basis can be obtained. For example, notice that the bubble coefficients of  any four point amplitudes are determined by the product of two four point tree level amplitudes. And since the four point amplitude basis of any dimension can be expressed in a uniform way: the product of spinor contractions and polynomials of Mandestam variables $s$ and $t$, the RG running of this kind of basis should also be able to expressed in the uniform form. Since in this work we focus on how to use on-shell method to derive the RG running in SMEFT, we just give a simple example to confirm this claim. More universal results will be presented in the future work. We will show the RG running for $H^2 B^+ B^- D^{2n}$ type amplitude basis at any dimensions generated from the general $H^4 D^{2n+4}$ type amplitude basis.  These basis can be expressed uniformly as
\bea
\mathcal{A}(H^\alpha  H^\beta  H^{\dagger \dot \alpha} H^{\dagger \dot \beta}) &\supset&  T^+ _{\alpha\beta\dot\alpha\dot\beta} C_{m,2n}^{H^4+} s_{12}^m (s_{13} -s_{23})^{2n}, \; \; T^- _{\alpha\beta\dot\alpha\dot\beta} C_{m,2n+1}^{H^4-} s_{12}^m (s_{13} -s_{23})^{2n+1}, \nonumber \\
\mathcal{A}( B^{+} B^{-}H^\alpha H^{\dagger \dot \beta}) &=&C_{m,n}^{H^2B^+B^-}[1|p_3 \ket{2}^2 s_{12}^m (s_{13} -s_{23})^{n}. 
\eea

Applying the on-shell method as above, we can obtain the universal RG running for $C_{m,n}^{H^2B^+B^-}$, compactly collected as,
\bea
\dot C_{m,n}^{H^2B^+B^-}&=&-4y_h^2 \Big(\sum^{i,j}_{i+2j=m+n+2}3C_{i,2j}^{H^4+}F(m,n,i,2j)+\sum^{i',j'}_{i'+2j'=m+n+1}C_{i',2j'+1}^{H^4-}F(m,n,i',2j'+1)\Big),\nonumber\\
F(m,n,i,k)&=&\sum^{g,f,h,l,d}_{2g+f+h=n} 
(\frac{1}{4})^{l-1}(-1)^gC_{i}^{2l-d}C_{2j}^{d}C_{l-1}^{g}C_{i-2l+d}^{h}C_{2l}^{l+1}C_{2j-d}^{l}\nonumber \\
&\times&\int_0^1 (1-\frac{1}{2t})^h(-\frac{1}{2t})^{m-2l+d-h}(-\frac{1}{2}+\frac{t}{2}+\frac{1}{2t})^f(\frac{3}{2}+\frac{t}{3}-\frac{1}{2t})^{2j-d-f}(\frac{1}{t}-1)^lt^{d+i} dt.
\eea
So the anomalous dimensions for $H^2 B^+ B^-D^{2n}$ basis at any dimension can be readily read from this universal expression. For example, the amplitude basis at dimension 8, or $(m=0,n=0)$, can get three contributions in above sum,
\bea
(i=0,j=1,d=2,l=1,g=h=f=0);\nonumber \\
(i=2,j=0,d=0,l=1,g=h=f=0);\nonumber \\
(i'=1,j'=0,d=1,l=1,g=h=f=0).
\eea
 With a simple integral $\int _0^1(\frac{1}{t}-1)t^2=\frac{1}{6}$,
we can obtain its RG running,
\bea
\dot C_{00}^{H^2B^+B^-}&=&-2g_1^2 y_h^2 (C_{0,2}^{H^4+}+C_{2,0}^{H^4+}+\frac{1}{3}C_{1,1}^{H^4-}),
\eea
which recovers the result we obtained in previous section.

\section{Conclusion}
\label{sec:conclusion}
The on-shell amplitude methods have remarkable advantages in studying SMEFT. The non-renormalizable interactions can be described by unfactorizable amplitude basis without worrying about redundancies. Comparing with Feynman diagrams calculations, the loop-level amplitudes can be constructed by unitary cut very efficiently, because there are no unphysical degrees of freedom in on-shell amplitudes. Especially, since the UV divergences of one-loop amplitudes are only from scalar bubble integrals and the bubble coefficients are related to the tree-level amplitudes, the on-shell method is very convenient to obtain the renormalization group running of higher dimension operators (amplitude basis). 

In this work we demonstrate in detail how to extract the full UV divergences of the loop amplitudes. The coefficients of massive bubble integrals can be extracted by Stokes' Theorem analytically. However, the UV divergences from massless bubble integrals are canceled by collinear IR divergences so they can not be obtained by unitary cut. However, they can be extracted from the collinear divergences of tree-level amplitudes. Since they are universal in SMEFT, we calculate the collinear divergence factors for all the standard model fields and list them in Eq.~\ref{eq:collinear}. So the UV divergences from the massless bubbles can be directly read from the list.

We present some examples to show how to derive the anomalous dimension matrix correctly. Some selection rules can be easily observed based on this method, like the ones from angular momentum conservation or global symmetries observed in our calculations for some dimension 8 operators. This method makes the structure of anomalous dimension matrix transparent, so the running of the general 4-point amplitude basis can be expressed in a universal form. We present the universal expressions for the anomalous dimension matrix of the general amplitude basis $H^2 B^+ B^- D^{2n}$ generated from the contributions of general $H^4 D^{2n+4}$ type amplitude basis at one-loop level. 

\section*{Note added}
While this paper was being finalized, Ref.~\cite{EliasMiro:2020tdv,Baratella:2020lzz} appeared, which presents a similar topic. \cite{EliasMiro:2020tdv} uses both form factors and on-shell amplitudes in their calculations and gives the anomalous dimensions at two loops. \cite{Baratella:2020lzz} uses the on-shell unitary cut similar to us but considers only the mixing between different operators, for which there are no IR contributions. In this paper, we take a pure on-shell method and demonstrate the complete procedure in deriving the anomalous dimension at one loop,  obtaining the bubble coefficients  and subtracting the collinear divergences. We also give new results in dimension 8 as well as universal expressions in general dimensions.

\acknowledgments
We thank Rui Yu, Qingjun Jin and Gang Yang for useful discussions. J.S. is supported by the National Natural Science Foundation of China (NSFC) under grant No.11947302, No.11690022, No.11851302, No.11675243 and No.11761141011 and also supported by the Strategic Priority Research Program of the Chinese Academy of Sciences under grant No.XDB21010200 and No.XDB23000000. T.M. are supported by the United States-Israel Binational Science Foundation~(BSF) (NSF-BSF program Grant No. 2018683) and the Azrieli foundation.
\appendix

\section{Example of deriving  the collinear factor for Higgs and Hypercharge gauge boson}
\label{collinear}
In this appendix, we show how to get the collinear divergence factor for Higgs in $U(1)_Y$ gauge  theory, following the procedure in ~\cite{Giele:1991vf,Giele:1993dj}. 
To get the collinear factor of Higgs leg generated from $U(1)_Y$ gauge interactions, we consider the scattering amplitude $HH^\dagger \to B^+ B^-$.  We suppose the Higgs leg $H$ with momentum $p_1$ and $B^+$ with momentum $p_3$  become collinear, which makes the Higgs propagator attached with these two legs on-shell and leads to divergence. We can parametrize the momenta as 
\bea
\ket{1}=\sqrt{z} \ket{P}, \quad \ket{3} =\sqrt{1-z} \ket{P},
\eea
where $0< z < 1$ and $P$ is the momentum of the Higgs propagator.     

In the collinear limit, the four point amplitudes $\mathcal{A}(H H^\dagger B^+ B^-)$ factorize into the product of  three point amplitude and a singular splitting function   
\bea
\mathcal{A}(H(p_1) H^\dagger (p_2) B^+(p_3) B^-(p_4)) \to  \text{Split}_{H}(H(p_1), B^+(p_3)) \mathcal{A}(H(P)  H^\dagger(p_2) B^-(p_4) ),
\eea
where the spliting function $\text{Split}_{H}(H(p_1), B^+(p_3)) =\sqrt{2}g_1 y_h \frac{\sqrt{z}}{\sqrt{1-z} \vev{13}}$, and the three point amplitude $ \mathcal{A}(H(P)  H^\dagger(p_2) B^-(p_4) ) =\sqrt{2}g_1 y_h  \frac{\vev {4P} \vev {42} }{\vev{P2}}$.   
The collinear factor $c_F^{HB \to H^\dagger}$ can be expressed as
\bea
c_F^{HB \to H^\dagger} = \sum_{i= \pm} \text{Split}_{H}(H(p_1), B^i(p_3))   \text{Split}_{H}^\dagger(H(p_1), B^i(p_3))     =   \frac{4 g_1^2 y_h^2}{\vev{13}[13]} \frac{z}{1-z}
\eea
Notice that the contributions from different polarizations of the gauge boson $B_\mu$ should be included. 
The phase space of these two collinear legs can be expressed as~\cite{Giele:1991vf}
\bea
dP^{\epsilon}_{col}(p_1,p_3,z) =\frac{(4\pi)^{\epsilon}}{16\pi^2\Gamma(1-\epsilon)} ds_{13} dz [s_{13} z(1-z)]^{-\epsilon}\theta(s_{min} -s_{13}).
\eea

Putting these together, we can get the collinear divergence of the process $HH^\dagger \to B^+ B^-$
\bea \label{eq:col}
\int c_F^{HB \to H^\dagger} dP^{\epsilon}_{col}(p_1,p_3,z) =\frac{g_1^2 y_h^2}{4\pi^2} \frac{1}.{\Gamma(1-\epsilon)} \frac{1}{\epsilon}.
\eea

The tree level collinear divergences should be cancelled by IR divergence from loop amplitudses. If we parametrize the one-loop collinear divergences as
\bea
\mathcal{A}_{\text{n,col}}^{\text{1-loop}} = -(\frac{1}{4\pi} )^2 \sum_{a}^n \frac{\gamma(a)}{\epsilon} \mathcal{A}^{\text{tree}},  
\eea
the collinear divergence factor for Higgs leg (only including the $U(1)_Y$ gauge corrections) can be extracted to be
\bea
\gamma(H) =2 g_1^2 y_h^2. 
\eea

Following the same procedure, we can extract the spliting function for vetex $H(p_1)H^\dagger (p_2) \to B$ from the process $H^a  H^{\dagger \dot{b}} \to H^c H^{\dagger \dot{d}}$, $ \text{Split}_{B^\pm}(H^a, H^{\dagger \dot{b}})  = \delta^{a \dot{b}} \sqrt{2} g_1 y_h  \sqrt{z (1-z)}/\vev{12}$. The collinear factor is given by    
\bea
c_F^{H(p_1)H^\dagger(p_2) \to B^\pm} = \sum_{a,\dot{b}}   \text{Split}_{B^\pm}(H^a, H^{\dagger \dot{b}})   \text{Split}_{B^\pm}^\dagger(H^a, H^{\dagger \dot{b}})    =   \frac{4 g_1^2 y_h^2}{\vev{12}[12]}z(1-z).
\eea
Do the same integration as in Eq.~\ref{eq:col}, we can easily get the collinear divergence factor for the $U(1)_Y$ gauge boson legs (only include the Higgs doublet corrections)
\bea
\gamma(B^\pm) =-\frac{g_1^2 y_h^2}{3}. 
\eea

\section{The anomalous dimension of $O_{eW}$}
\label{app:Oew}
In this appendix we present the full calculation of the running of $O_{eW}=(\bar{l}\sigma^{\mu \nu} e)\tau^IH W^I_{\mu \nu}$. The contribution from different unitary cuts are summarized in table.\ref{tab:cutO_ew}, where $\mathcal{A}^6$ and $\mathcal{A}^{\text{SM}}$ are dimension 6 and SM on-shell amplitudes at the two sides of the cut respectively. The relevant form of dimension 6 operators and expression of SM amplitudes are summarized in table.\ref{tab:ops&amps}

Combined with the contributions from collinear divergences read from Eq.\ref{eq:collinear}:
\bea
\dot{C}_{eW}=-2(\gamma(W)+\gamma(H)+\gamma(l)+\gamma(e))C_{eW},
\eea
we obtain the same result as that calculated from Feynman diagrams\cite{Jenkins:2013wua,Alonso:2013hga,Jenkins:2013zja}\footnote{There is a relative minus signs for terms linear in $g_i$ due to conventions.}.
\begin{table}[htbp]
    \centering
 {\tiny   \begin{tabular}{c|c|c}
		\hline\hline
		$\mathcal{A}^6$	&	$\mathcal{A^{\text{SM}}}	$			&	$\text{Contribution to } \dot{C}_{eW}	$\\
		\hline
		$\mathcal{A}(W^{i-}W^{i-}H^{\dagger \dot \alpha}H^\alpha)=2(C_{HW}+iC_{H\tilde{W}})\delta^{\alpha \dot \alpha}\l12\r^2$	&	$\mathcal{A}(l^{\dot \alpha -}e^-W^{i+}H^\beta)$
	&	$Y_e^\dagger g_2 (C_{HW}+i C_{H\widetilde{W}})$\\
	\hline
		$\mathcal{A}(W^{i-}B^-H^{\dagger \dot \alpha}H^{\beta})(C_{HWB}+i C_{H\widetilde{W}B})(\tau^i)^{\dot \alpha}_\beta\l12\r^2$	&$	\mathcal{A}(l^{\dot \alpha -}e^-B^{+}H^\alpha)$	&	$Y_e^\dagger (y_e+y_l) g_1 (C_{HWB}+i C_{H\widetilde{W}B})$\\
		\hline
		$\mathcal{A}(l_{\dot \alpha}^-e^-q_{\beta}^-u^-)=[(C^1_{lequ}-4C^3_{lequ})\l12\r\l34\r-8C^3_{lequ}\l14\r\l32\r]\epsilon_{\dot \alpha \beta}$&	$\mathcal{A}^{SM}(u^+q^{-}_{\beta}W^{i-}H_{\sigma})$&$	2Y_u N_c g_2 C^3_{lequ}$\\
				\hline
			$\mathcal{A}(l^{\dot \alpha -}e^-B^{-}H^\alpha)=-2\sqrt{2}C_{eB}\delta^{\alpha \dot \alpha}\l 13 \r \l23\r	$&	$\mathcal{A}(H^\beta H^{\dagger\dot \alpha}B^+W^{i-})	$&	$2y_hg_1 g_2 C_{eB}$\\
		\cline{2-3}
		 &$\mathcal{A}(l^{+\beta}l^{-\dot \alpha}B^+W^{i-})$ & $y_l g_1 g_2 C_{eB}$\\ 
		\hline
		$\mathcal{A}(l^{\dot \alpha -}e^-W^{i-}H^\beta)=-2\sqrt{2}C_{eW}(\tau^i)^{\dot \alpha}_\beta \l 13 \r \l23\r$	&	$\mathcal{A}(H^{\dagger \dot \alpha}H^\alpha e^+e^-)$ &$ Y_e^\dagger Y_e C_{eW}$\\
		\cline{2-3}
           &$\mathcal{A}(l^{-\dot \alpha}e^-l^{+\alpha}e^+)$ & $4g_1^2y_h y_e$\\
           \cline{2-3}
         &  $\mathcal{A}(e^{-}e^{+}H^{\dagger \sigma}H^{\beta})$&         $- 4 g_1^2y_h y_e$\\
            \cline{2-3}
         & $\mathcal{A}(l^{-\dot \alpha}l^{+\gamma}H^{\dagger \sigma}H^{\beta}) $& $4\big(g_2^2(C_2(2)-\frac{1}{2}C_2(G))+g'^2y_h y_l\big)$\\
                \cline{2-3}
          & $\mathcal{A}(H^{\dagger \dot \alpha}W^{j+}W^{i-}H^{\beta})$ & $2g^2(C_2(2)+C_2(G))$\\
                 \cline{2-3}
          &$\mathcal{A}(l^{-\dot \alpha}l^{+\beta}W^{i-}W^{j+})$ & $4g^2C_2(2)$\\
		\hline\hline
\end{tabular}}
\caption{Contributions to the running of $O_{eW}$  from different unitary cuts.  $\mathcal{A}^6$ are on-shell amplitudes from dimension 6 operators and $\mathcal{A}^{SM}$  are SM amplitudes. Here $t^j t^it^j=(C_2(2)-\frac{1}{2}C_2(G))t^i$.}\label{tab:cutO_ew}
\end{table}

\begin{table}[htbp]
    \centering
    \begin{tabular}{lc|}
\hline \hline \vspace{-0.4cm} \\
$O_{HW}=H^\dagger H W^a_{\mu \nu }W^{a\mu\nu}$ \\
$O_{H\tilde{W}}=H^\dagger  H \widetilde{W}^a_{\mu \nu }W^{a\mu\nu} $\\
$O_{HWB}=H^\dagger \tau ^a H W^a_{\mu \nu} B^{\mu \nu}$ \\ 
$O_{H\tilde{W}B}=H^\dagger \tau ^a H \widetilde{W} ^a_{\mu \nu} B^{\mu \nu}$ \\
$O^1_{lequ}=(\bar{l}^\alpha e)\epsilon_{\alpha \beta} (\bar{q}^\beta e)$ \\
$O^3_{lequ}=(\bar{l}^\alpha\sigma_{\mu \nu} e)\epsilon_{\alpha \beta} (\bar{q}^\beta \sigma^{\mu \nu} e)$\\
$O_{eB}=(\bar{l}\sigma^{\mu \nu} e)H B_{\mu \nu}$ \\
$O_{eW}=(\bar{l}\sigma^{\mu \nu} e)\tau^aH W^a_{\mu \nu}$\\ 
\vspace{-0.4cm} \\
\hline \hline
\end{tabular}
 {\small   \begin{tabular}{|c|c|}
	\hline
		$\mathcal{A}(l^{\dot \alpha -}e^-W^{i+}H^\beta)$&$-\sqrt{2} Y_e^\dagger g_2 \frac{(\tau^i)^{\dot\alpha}_\beta}{2}\frac{\l12\r}{[12]}\frac{[23]}{\l13\r}$
	\\
	\hline
			$\mathcal{A}(l^{\dot \alpha -}e^-B^{+}H^\alpha)$&$-\sqrt{2}\delta^{\dot \alpha}_{\alpha} Y_e^\dagger g_1 \frac{\l12\r}{[12]}(y_e \frac{[13]}{\l23\r}+y_l \frac{[23]}{\l13\r})$	\\
		\hline
				$\mathcal{A}(u^+q^{-}_{\beta}W^{i-}H_{\sigma})$&$\sqrt{2} Y_u g_2 \epsilon_{\lambda \sigma} \frac{(\tau^i)^\lambda_\beta}{2}\frac{\l13\r [21]}{[23]\l12\r}	$\\
		\hline
		$\mathcal{A}(H^\beta H^{\dot \dagger\alpha}B^+W^{i-})$&$-g_1g_2y_h(\tau^i)^{\dot\alpha} _\beta \frac{\vev{14} \vev{24}}{\vev{13} \vev{23}}	$\\
		\hline
		 $\mathcal{A}(l^{+\beta}l^{\dot\alpha-}B^+W^{i-})$&$g_1g_2y_l(\tau^i)^{\dot\alpha} _\beta \frac{\vev{24}^2}{\vev{13} \vev{23}}$ \\ 
		\hline
			$\mathcal{A}(H^{\dagger \dot \alpha}H^\alpha e^+e^-)$&$-Y_e^\dagger Y_e \frac{\l14\r}{\l13\r}$ \\
		\hline
           $\mathcal{A}(l^{\dot \alpha -}e^-l^{+\alpha}e^+)$&$2y_l y_e g_1^2\delta^{\dot \alpha}_\alpha\frac{\l12\r [34]}{\l13\r [13]}$ \\
           \hline
         $\mathcal{A}(e^{-}e^{+}H^{\dagger \dot \sigma}H^{\beta})$&$2y_h y_e g_1^2 \delta^{\dot \sigma}_\beta\frac{\l1|4|2]}{\l12\r[12]}$\\
            \hline
         $\mathcal{A}(l^{\dot \alpha -}l^{+\gamma}H^{\dagger \dot \sigma}H^{\beta})$&$-2\big(y_h y_l g_1^2 \delta^{\dot\alpha}_\gamma \delta^{\dot \sigma}_\beta +g_2^2(t^i)^{\dot\alpha}_\gamma(t^i)^{\dot \sigma}_\beta \big)\frac{\l1|4|2]}{\l12\r[12]} $\\
                \hline
           $\mathcal{A}(H^{\dagger \dot \alpha}W^{j+}W^{i-}H^{\beta})$&$-2g_2^2\frac{\l3|1|2]^2}{\l23\r[23]}\big(\frac{(t^it^j)^{\dot\alpha}_\beta}{\l13\r[13]}+\frac{(t^j t^i)^{\dot\alpha}_\beta}{\l12\r[12]}\big)$ \\
                 \hline
          $\mathcal{A}(l^{\dot\alpha-}l^{+\beta}W^{i-}W^{j+})$&$2g_2^2\frac{\l13\r [24]}{\l34\r[43]}\big( \frac{\l13\r}{\l14\r}(t^j t^i)^{\dot\alpha}_\beta+\frac{[14]}{[13]}(t^i t^j)^{\dot\alpha}_\beta \big)$ \\
		\hline
\end{tabular}}
\caption{Left: Dimension 6 operators that contribute to the running of $O_{eW}$; Right: Expressions of SM amplitudes used in calculating $\dot{C}_{eW}$.}\label{tab:ops&amps}
\end{table}




\bibliography{RG}

\providecommand{\href}[2]{#2}\begingroup\raggedright\begin{thebibliography}{10}

\bibitem{Chatrchyan:2012ufa}
{\scshape CMS} collaboration, \emph{{Observation of a New Boson at a Mass of
  125 GeV with the CMS Experiment at the LHC}},
  \href{https://doi.org/10.1016/j.physletb.2012.08.021}{\emph{Phys. Lett. B}
  {\bfseries 716} (2012) 30} [\href{https://arxiv.org/abs/1207.7235}{{\ttfamily
  1207.7235}}].

\bibitem{Aad:2012tfa}
{\scshape ATLAS} collaboration, \emph{{Observation of a new particle in the
  search for the Standard Model Higgs boson with the ATLAS detector at the
  LHC}}, \href{https://doi.org/10.1016/j.physletb.2012.08.020}{\emph{Phys.
  Lett. B} {\bfseries 716} (2012) 1}
  [\href{https://arxiv.org/abs/1207.7214}{{\ttfamily 1207.7214}}].

\bibitem{Wess:1974tw}
J.~Wess and B.~Zumino, \emph{{Supergauge Transformations in Four-Dimensions}},
  \href{https://doi.org/10.1016/0550-3213(74)90355-1}{\emph{Nucl. Phys. B}
  {\bfseries 70} (1974) 39}.

\bibitem{Volkov:1973ix}
D.~Volkov and V.~Akulov, \emph{{Is the Neutrino a Goldstone Particle?}},
  \href{https://doi.org/10.1016/0370-2693(73)90490-5}{\emph{Phys. Lett. B}
  {\bfseries 46} (1973) 109}.

\bibitem{Kaplan:1983fs}
D.~B. Kaplan and H.~Georgi, \emph{{SU(2) x U(1) Breaking by Vacuum
  Misalignment}},
  \href{https://doi.org/10.1016/0370-2693(84)91177-8}{\emph{Phys. Lett. B}
  {\bfseries 136} (1984) 183}.

\bibitem{Georgi:1984af}
H.~Georgi and D.~B. Kaplan, \emph{{Composite Higgs and Custodial SU(2)}},
  \href{https://doi.org/10.1016/0370-2693(84)90341-1}{\emph{Phys. Lett. B}
  {\bfseries 145} (1984) 216}.

\bibitem{Dugan:1984hq}
M.~J. Dugan, H.~Georgi and D.~B. Kaplan, \emph{{Anatomy of a Composite Higgs
  Model}}, \href{https://doi.org/10.1016/0550-3213(85)90221-4}{\emph{Nucl.
  Phys. B} {\bfseries 254} (1985) 299}.

\bibitem{Shadmi:2018xan}
Y.~Shadmi and Y.~Weiss, \emph{{Effective Field Theory Amplitudes the On-Shell
  Way: Scalar and Vector Couplings to Gluons}},
  \href{https://doi.org/10.1007/JHEP02(2019)165}{\emph{JHEP} {\bfseries 02}
  (2019) 165} [\href{https://arxiv.org/abs/1809.09644}{{\ttfamily
  1809.09644}}].

\bibitem{Ma:2019gtx}
T.~Ma, J.~Shu and M.-L. Xiao, \emph{{Standard Model Effective Field Theory from
  On-shell Amplitudes}},  \href{https://arxiv.org/abs/1902.06752}{{\ttfamily
  1902.06752}}.

\bibitem{Cachazo:2013iea}
F.~Cachazo, S.~He and E.~Y. Yuan, \emph{{Scattering of Massless Particles:
  Scalars, Gluons and Gravitons}},
  \href{https://doi.org/10.1007/JHEP07(2014)033}{\emph{JHEP} {\bfseries 07}
  (2014) 033} [\href{https://arxiv.org/abs/1309.0885}{{\ttfamily 1309.0885}}].

\bibitem{ArkaniHamed:2008gz}
N.~Arkani-Hamed, F.~Cachazo and J.~Kaplan, \emph{{What is the Simplest Quantum
  Field Theory?}}, \href{https://doi.org/10.1007/JHEP09(2010)016}{\emph{JHEP}
  {\bfseries 09} (2010) 016} [\href{https://arxiv.org/abs/0808.1446}{{\ttfamily
  0808.1446}}].

\bibitem{Cheung:2016drk}
C.~Cheung, K.~Kampf, J.~Novotny, C.-H. Shen and J.~Trnka, \emph{{A Periodic
  Table of Effective Field Theories}},
  \href{https://doi.org/10.1007/JHEP02(2017)020}{\emph{JHEP} {\bfseries 02}
  (2017) 020} [\href{https://arxiv.org/abs/1611.03137}{{\ttfamily
  1611.03137}}].

\bibitem{Low:2014nga}
I.~Low, \emph{{Adler's zero and effective Lagrangians for nonlinearly realized
  symmetry}}, \href{https://doi.org/10.1103/PhysRevD.91.105017}{\emph{Phys.
  Rev. D} {\bfseries 91} (2015) 105017}
  [\href{https://arxiv.org/abs/1412.2145}{{\ttfamily 1412.2145}}].

\bibitem{Cheung:2015aba}
C.~Cheung and C.-H. Shen, \emph{{Nonrenormalization Theorems without
  Supersymmetry}},
  \href{https://doi.org/10.1103/PhysRevLett.115.071601}{\emph{Phys. Rev. Lett.}
  {\bfseries 115} (2015) 071601}
  [\href{https://arxiv.org/abs/1505.01844}{{\ttfamily 1505.01844}}].

\bibitem{Bern:2019wie}
Z.~Bern, J.~Parra-Martinez and E.~Sawyer, \emph{{Nonrenormalization and
  Operator Mixing via On-Shell Methods}},
  \href{https://doi.org/10.1103/PhysRevLett.124.051601}{\emph{Phys. Rev. Lett.}
  {\bfseries 124} (2020) 051601}
  [\href{https://arxiv.org/abs/1910.05831}{{\ttfamily 1910.05831}}].

\bibitem{Jiang:2020sdh}
M.~Jiang, J.~Shu, M.-L. Xiao and Y.-H. Zheng, \emph{{New Selection Rules from
  Angular Momentum Conservation}},
  \href{https://arxiv.org/abs/2001.04481}{{\ttfamily 2001.04481}}.

\bibitem{Bern:1994zx}
Z.~Bern, L.~J. Dixon, D.~C. Dunbar and D.~A. Kosower, \emph{{One loop n point
  gauge theory amplitudes, unitarity and collinear limits}},
  \href{https://doi.org/10.1016/0550-3213(94)90179-1}{\emph{Nucl. Phys. B}
  {\bfseries 425} (1994) 217}
  [\href{https://arxiv.org/abs/hep-ph/9403226}{{\ttfamily hep-ph/9403226}}].

\bibitem{Bern:1994cg}
Z.~Bern, L.~J. Dixon, D.~C. Dunbar and D.~A. Kosower, \emph{{Fusing gauge
  theory tree amplitudes into loop amplitudes}},
  \href{https://doi.org/10.1016/0550-3213(94)00488-Z}{\emph{Nucl. Phys. B}
  {\bfseries 435} (1995) 59}
  [\href{https://arxiv.org/abs/hep-ph/9409265}{{\ttfamily hep-ph/9409265}}].

\bibitem{Bern:1997sc}
Z.~Bern, L.~J. Dixon and D.~A. Kosower, \emph{{One loop amplitudes for e+ e- to
  four partons}},
  \href{https://doi.org/10.1016/S0550-3213(97)00703-7}{\emph{Nucl. Phys. B}
  {\bfseries 513} (1998) 3}
  [\href{https://arxiv.org/abs/hep-ph/9708239}{{\ttfamily hep-ph/9708239}}].

\bibitem{Mastrolia:2009dr}
P.~Mastrolia, \emph{{Double-Cut of Scattering Amplitudes and Stokes' Theorem}},
  \href{https://doi.org/10.1016/j.physletb.2009.06.033}{\emph{Phys. Lett. B}
  {\bfseries 678} (2009) 246}
  [\href{https://arxiv.org/abs/0905.2909}{{\ttfamily 0905.2909}}].

\bibitem{Huang:2012aq}
Y.-t. Huang, D.~A. McGady and C.~Peng, \emph{{One-loop renormalization and the
  S-matrix}}, \href{https://doi.org/10.1103/PhysRevD.87.085028}{\emph{Phys.
  Rev. D} {\bfseries 87} (2013) 085028}
  [\href{https://arxiv.org/abs/1205.5606}{{\ttfamily 1205.5606}}].

\bibitem{Forde:2007mi}
D.~Forde, \emph{{Direct extraction of one-loop integral coefficients}},
  \href{https://doi.org/10.1103/PhysRevD.75.125019}{\emph{Phys. Rev. D}
  {\bfseries 75} (2007) 125019}
  [\href{https://arxiv.org/abs/0704.1835}{{\ttfamily 0704.1835}}].

\bibitem{Anastasiou:2006jv}
C.~Anastasiou, R.~Britto, B.~Feng, Z.~Kunszt and P.~Mastrolia,
  \emph{{D-dimensional unitarity cut method}},
  \href{https://doi.org/10.1016/j.physletb.2006.12.022}{\emph{Phys. Lett. B}
  {\bfseries 645} (2007) 213}
  [\href{https://arxiv.org/abs/hep-ph/0609191}{{\ttfamily hep-ph/0609191}}].

\bibitem{Giele:1991vf}
W.~Giele and E.~Glover, \emph{{Higher order corrections to jet cross-sections
  in e+ e- annihilation}},
  \href{https://doi.org/10.1103/PhysRevD.46.1980}{\emph{Phys. Rev. D}
  {\bfseries 46} (1992) 1980}.

\bibitem{Giele:1993dj}
W.~Giele, E.~Glover and D.~A. Kosower, \emph{{Higher order corrections to jet
  cross-sections in hadron colliders}},
  \href{https://doi.org/10.1016/0550-3213(93)90365-V}{\emph{Nucl. Phys. B}
  {\bfseries 403} (1993) 633}
  [\href{https://arxiv.org/abs/hep-ph/9302225}{{\ttfamily hep-ph/9302225}}].

\bibitem{Kunszt:1994np}
Z.~Kunszt, A.~Signer and Z.~Trocsanyi, \emph{{Singular terms of helicity
  amplitudes at one loop in QCD and the soft limit of the cross-sections of
  multiparton processes}},
  \href{https://doi.org/10.1016/0550-3213(94)90077-9}{\emph{Nucl. Phys. B}
  {\bfseries 420} (1994) 550}
  [\href{https://arxiv.org/abs/hep-ph/9401294}{{\ttfamily hep-ph/9401294}}].

\bibitem{Grojean:2013kd}
C.~Grojean, E.~E. Jenkins, A.~V. Manohar and M.~Trott, \emph{{Renormalization
  Group Scaling of Higgs Operators and $\Gamma(h \rightarrow \gamma \gamma)$}},
  \href{https://doi.org/10.1007/JHEP04(2013)016}{\emph{JHEP} {\bfseries 04}
  (2013) 016} [\href{https://arxiv.org/abs/1301.2588}{{\ttfamily 1301.2588}}].

\bibitem{Jenkins:2013wua}
E.~E. Jenkins, A.~V. Manohar and M.~Trott, \emph{{Renormalization Group
  Evolution of the Standard Model Dimension Six Operators II: Yukawa
  Dependence}}, \href{https://doi.org/10.1007/JHEP01(2014)035}{\emph{JHEP}
  {\bfseries 01} (2014) 035} [\href{https://arxiv.org/abs/1310.4838}{{\ttfamily
  1310.4838}}].

\bibitem{Alonso:2013hga}
R.~Alonso, E.~E. Jenkins, A.~V. Manohar and M.~Trott, \emph{{Renormalization
  Group Evolution of the Standard Model Dimension Six Operators III: Gauge
  Coupling Dependence and Phenomenology}},
  \href{https://doi.org/10.1007/JHEP04(2014)159}{\emph{JHEP} {\bfseries 04}
  (2014) 159} [\href{https://arxiv.org/abs/1312.2014}{{\ttfamily 1312.2014}}].

\bibitem{Jenkins:2013zja}
E.~E. Jenkins, A.~V. Manohar and M.~Trott, \emph{{Renormalization Group
  Evolution of the Standard Model Dimension Six Operators I: Formalism and
  lambda Dependence}},
  \href{https://doi.org/10.1007/JHEP10(2013)087}{\emph{JHEP} {\bfseries 10}
  (2013) 087} [\href{https://arxiv.org/abs/1308.2627}{{\ttfamily 1308.2627}}].

\bibitem{Alonso:2014zka}
R.~Alonso, H.-M. Chang, E.~E. Jenkins, A.~V. Manohar and B.~Shotwell,
  \emph{{Renormalization group evolution of dimension-six baryon number
  violating operators}},
  \href{https://doi.org/10.1016/j.physletb.2014.05.065}{\emph{Phys. Lett. B}
  {\bfseries 734} (2014) 302}
  [\href{https://arxiv.org/abs/1405.0486}{{\ttfamily 1405.0486}}].

\bibitem{Li:2020gnx}
H.-L. Li, Z.~Ren, J.~Shu, M.-L. Xiao, J.-H. Yu and Y.-H. Zheng, \emph{{Complete
  Set of Dimension-8 Operators in the Standard Model Effective Field Theory}},
  \href{https://arxiv.org/abs/2005.00008}{{\ttfamily 2005.00008}}.

\bibitem{Murphy:2020rsh}
C.~W. Murphy, \emph{{Dimension-8 Operators in the Standard Model Effective
  Field Theory}},  \href{https://arxiv.org/abs/2005.00059}{{\ttfamily
  2005.00059}}.

\bibitem{Bern:2004cz}
Z.~Bern, L.~J. Dixon and D.~A. Kosower, \emph{{Two-loop $g\rightarrow gg$
  splitting amplitudes in QCD}},
  \href{https://doi.org/10.1088/1126-6708/2004/08/012}{\emph{JHEP} {\bfseries
  08} (2004) 012} [\href{https://arxiv.org/abs/hep-ph/0404293}{{\ttfamily
  hep-ph/0404293}}].

\bibitem{Bern:2004ky}
Z.~Bern, V.~Del~Duca, L.~J. Dixon and D.~A. Kosower, \emph{{All
  non-maximally-helicity-violating one-loop seven-gluon amplitudes in N=4
  super-yang-Mills theory}},
  \href{https://doi.org/10.1103/PhysRevD.71.045006}{\emph{Phys. Rev. D}
  {\bfseries 71} (2005) 045006}
  [\href{https://arxiv.org/abs/hep-th/0410224}{{\ttfamily hep-th/0410224}}].

\bibitem{Britto:2004nc}
R.~Britto, F.~Cachazo and B.~Feng, \emph{{Generalized unitarity and one-loop
  amplitudes in N=4 super-Yang-Mills}},
  \href{https://doi.org/10.1016/j.nuclphysb.2005.07.014}{\emph{Nucl. Phys. B}
  {\bfseries 725} (2005) 275}
  [\href{https://arxiv.org/abs/hep-th/0412103}{{\ttfamily hep-th/0412103}}].

\bibitem{Cohen:2010mi}
T.~Cohen, H.~Elvang and M.~Kiermaier, \emph{{On-shell constructibility of tree
  amplitudes in general field theories}},
  \href{https://doi.org/10.1007/JHEP04(2011)053}{\emph{JHEP} {\bfseries 04}
  (2011) 053} [\href{https://arxiv.org/abs/1010.0257}{{\ttfamily 1010.0257}}].

\bibitem{Azatov:2016sqh}
A.~Azatov, R.~Contino, C.~S. Machado and F.~Riva, \emph{{Helicity selection
  rules and noninterference for BSM amplitudes}},
  \href{https://doi.org/10.1103/PhysRevD.95.065014}{\emph{Phys. Rev. D}
  {\bfseries 95} (2017) 065014}
  [\href{https://arxiv.org/abs/1607.05236}{{\ttfamily 1607.05236}}].

\bibitem{Craig:2019wmo}
N.~Craig, M.~Jiang, Y.-Y. Li and D.~Sutherland, \emph{{Loops and Trees in
  Generic EFTs}},  \href{https://arxiv.org/abs/2001.00017}{{\ttfamily
  2001.00017}}.

\bibitem{Arzt:1993gz}
C.~Arzt, \emph{{Reduced effective Lagrangians}},
  \href{https://doi.org/10.1016/0370-2693(94)01419-D}{\emph{Phys. Lett.}
  {\bfseries B342} (1995) 189}
  [\href{https://arxiv.org/abs/hep-ph/9304230}{{\ttfamily hep-ph/9304230}}].

\bibitem{Giudice:2007fh}
G.~F. Giudice, C.~Grojean, A.~Pomarol and R.~Rattazzi, \emph{{The
  Strongly-Interacting Light Higgs}},
  \href{https://doi.org/10.1088/1126-6708/2007/06/045}{\emph{JHEP} {\bfseries
  06} (2007) 045} [\href{https://arxiv.org/abs/hep-ph/0703164}{{\ttfamily
  hep-ph/0703164}}].

\bibitem{Einhorn:2013kja}
M.~B. Einhorn and J.~Wudka, \emph{{The Bases of Effective Field Theories}},
  \href{https://doi.org/10.1016/j.nuclphysb.2013.08.023}{\emph{Nucl. Phys.}
  {\bfseries B876} (2013) 556}
  [\href{https://arxiv.org/abs/1307.0478}{{\ttfamily 1307.0478}}].

\bibitem{EliasMiro:2020tdv}
J.~Elias~Miro, J.~Ingoldby and M.~Riembau, \emph{{EFT anomalous dimensions from
  the S-matrix}},  \href{https://arxiv.org/abs/2005.06983}{{\ttfamily
  2005.06983}}.

\bibitem{Baratella:2020lzz}
P.~Baratella, C.~Fernandez and A.~Pomarol, \emph{{Renormalization of
  Higher-Dimensional Operators from On-shell Amplitudes}},
  \href{https://arxiv.org/abs/2005.07129}{{\ttfamily 2005.07129}}.

\end{thebibliography}\endgroup

\end{document}